\documentclass[fleqn,usenatbib]{mnras}

\usepackage{newtxtext,newtxmath}
\usepackage[T1]{fontenc}

\DeclareRobustCommand{\VAN}[3]{#2}
\let\VANthebibliography\thebibliography
\def\thebibliography{\DeclareRobustCommand{\VAN}[3]{##3}\VANthebibliography}

\usepackage{float}
\usepackage{subfig}

\usepackage{graphicx}	% Including figure files
\usepackage{amsmath}	% Advanced maths commands
\title[Intrinsic Alignment of Galaxy Clusters]{Galaxy clusters as intrinsic alignment tracers: present and future}

\author[C.J.G. Vedder \& N.E. Chisari]{
C. J. G. Vedder,$^{1}$\thanks{E-mail: c.j.g.vedder@students.uu.nl}
N. E. Chisari$^{1}$\thanks{E-mail: n.e.chisari@uu.nl}
\\
$^{1}$Institute for Theoretical Physics, Utrecht University, Princetonplein 5, 3584 CC Utrecht, The Netherlands.
}

\date{Accepted XXX. Received YYY; in original form ZZZ}

\pubyear{2020}

\begin{document}
\label{firstpage}
\pagerange{\pageref{firstpage}--\pageref{lastpage}}
\maketitle

\begin{abstract}
    Galaxies and clusters embedded in the large-scale structure of the Universe are observed to align in preferential directions. Galaxy alignment has been established as a potential probe for cosmological information, but the application of cluster alignments for these purposes remains unexplored. Clusters are observed to have a higher alignment amplitude than galaxies, but because galaxies are much more numerous, the trade-off in detectability between the two signals remains unclear. We present forecasts comparing cluster and galaxy alignments for two extragalactic survey set-ups: a currently-available low redshift survey (SDSS) and an upcoming higher redshift survey (LSST). For SDSS, we rely on the publicly available {\tt redMaPPer} catalogue to describe the cluster sample. For LSST, we perform estimations of the expected number counts while we extrapolate the alignment measurements from SDSS. Clusters in SDSS have typically higher alignment signal-to-noise than galaxies. For LSST, the cluster alignment signals quickly wash out with redshift due to a relatively low number count and a decreasing alignment amplitude. Nevertheless, a potential strong-suit of clusters is in their interplay with weak lensing: intrinsic alignments can be more easily isolated for clusters than for galaxies. The signal-to-noise of cluster alignment can in general be improved by isolating close pairs along the line of sight.
\end{abstract}


%%%%%%%%%%%%%%%%%%%%%%%%%%%%%%%%%%%%%%%%%%%%%%%%%%

%%%%%%%%%%%%%%%%% BODY OF PAPER %%%%%%%%%%%%%%%%%%

\section{Introduction}

Next generation galaxy surveys such as \textit{Euclid} \citep{laureijs2011euclid}, the \textit{Nancy Roman Telescope} \citep{spergel2015widefield}, and the Legacy Survey of Space and Time -LSST- \citep{lsst}, will deliver accurate shape measurements for an unprecedented number of galaxies. These shapes contain a wealth of cosmological information \citep{van_Uitert_2018, Hildebrandt_2016, Joudaki_2020, johnston_2019,Samuroff_2019, Schmidt_2015, Chisari_2013, shapes, Chisari_2014}. The most studied one is gravitational lensing: the effect of photons being deflected during their trajectory to Earth, thereby altering the observed shape of the galaxy. In the weak regime, gravitational lensing is measured statistically from large samples. (For reviews on weak gravitational lensing, see \citet{Kilbinger_2015} and \citet{Mandelbaum}.) 

However, to interpret the measured statistics of galaxy shapes correctly one also has to account for intrinsic correlations between them. Galaxies are subject to interactions with the tidal field of the large-scale structure: when a structure collapses due to gravity, it does so along a preferential axis of the tidal field. On top of this, matter tends to be accreted onto a halo through this same preferential direction. This can result in a significant shape correlation between galaxies \citep{Catelan_2001,Hirata_2004}. Failing to account for this can yield significant biases in weak lensing  analyses \citep{Kirk_2012,Krause_2015}. Nevertheless, intrinsic galaxy alignments also contain cosmological information \citep{Chisari_2013, shapes, Chisari_2014, Schmidt_2015}. (For reviews on intrinsic alignments, see \citet{Troxel_2015} and \citet{Joachimi}.)

For Luminous Red Galaxies (LRGs) up to $z\sim 1$, these intrinsic alignments have been observed by several surveys \citep{Brown_2002,Joachimi_2011, Hirata_2007,Singh_2015,Blazek_2011,johnston_2019,Samuroff_2019} and they are also predicted by cosmological simulations \citep{Chisari_2015,Chisari_2016,Chisari_2017,Codis_2015, Croft_2000, Velliscig_2015, tenneti2020groupscale, Hilbert_2017,Bate_2019,Bhowmick_2019}. The strength of the alignment has been found to correlate with the luminosity of the galaxy \citep{Joachimi_2011,Singh_2015}. 

Galaxy clusters are groups of galaxies bound together by gravity and are therefore forming a distinct entity. Galaxy clusters can also be assigned a shape by determining the locations of member galaxies. This is generally found to trace the shape of the dark matter halo \citep{haloshape1,haloshape2} and to align with the large-scale structure \citep{Vanuitert_2017}. This alignment is stronger than for galaxies, and it follows a universal scaling with halo mass, both for clusters and for galaxies \citep{Vanuitert_2017, Piras_2017}. This suggests that the possibility of extracting cosmological information from cluster alignments should also be investigated. 

Galaxies, however, are much more numerous in the universe and although they align more weakly, the noise in their alignment measurement is smaller than for clusters. In this work, we investigate the trade-off between galaxy and cluster alignments in several survey set-ups. We make predictions for a next generation galaxy survey, LSST, by forecasting the properties of the cluster sample we expect to observe. We compare our results to the alignments of clusters \citep{Rykoff_2014} and galaxies from already observed samples by the Sloan Digital Sky Survey (SDSS, \citealt{sdssfull}). 

This work is organized as follows. First, we describe the observables and the model, galaxy and cluster samples and the method for forecasting the expected uncertainties in section \ref{sec:methods}. We present the results in section \ref{sec:results}. In section \ref{lensing}, we briefly discuss the role of weak gravitational lensing as a contaminant to alignment measurements of clusters. We conclude with the discussion and conclusion in sections \ref{sec:discussion} and \ref{sec:conclusion}, respectively.

Throughout this work we assume the following flat \textit{Planck} $\Lambda$CDM cosmology \citep{planck}: $\Omega_b=0.045$, $\Omega_{\rm CDM} = 0.27$, $h = H_0/(100\,{\rm km\,s^{-1} Mpc^{-1}})=0.67$, $\sigma_8 = 0.83$ and $n_s = 0.96$. Here, $\Omega_b$ and $\Omega_{\rm CDM}$ are the fractional energy densities of baryonic and cold dark matter, $h$ is the dimensionless Hubble constant, $\sigma_8$ is typical variance of density fluctuations on scales of $8 h^{-1}\,\operatorname{Mpc}$, and $n_s$ is the exponent of the power law that describes the power spectrum of the primordial potential during inflation. 

\section{Forecasting method and samples}
\label{sec:methods}

The shape of a galaxy or galaxy cluster is a sum of three components:
\begin{equation}\label{eq:shapesum}
    \gamma = \gamma^{IA} + \gamma^{WL} + \gamma^{\operatorname{rnd}}
\end{equation}
The first contribution is the intrinsic part - the observed shape due to intrinsic alignments with the large-scale structure; the second part is the contribution due to weak gravitational lensing; the last term is intrinsically random noise. 

\subsection{Intrinsic Alignment Model}
\label{sec:alignmentmodel} 

In the linear alignment model, the intrinsic shape of a galaxy or cluster is related to the primordial gravitational potential, $\phi_p(\vec{x})$, in the following way \citep{Catelan_2001,Hirata_2004}:
\begin{equation}\label{eq:gla}
    (\gamma_+^{IA}, \gamma_\times^{IA}) = -A_{IA} \frac{C_1 }{4\pi G}(\partial_x^2 - \partial_y^2,  2 \partial_x \partial_y)\phi_p(\vec{x}).
\end{equation}
By convention, $A_{IA}$ gives the amplitude of the signal in terms of a previously measured value by \citet{Brown_2002}, $C_1$. It is convenient to Fourier transform the expression in Eq. \ref{eq:gla} and write it in terms of an over-density using the Poisson equation. We then obtain, for example,
\begin{equation}\label{eq:ia}
        \gamma_+^{IA}(\vec{k}) = -A_{IA} \frac{C_1 \rho_{\rm crit} \Omega_{\rm M} }{D(z)}\frac{(k_x^2 - k_y^2)}{k^2}\delta(\vec{k}).
\end{equation}
Here, $D(z)$ is the growth function, $\Omega_{\rm M}$ is energy density in matter today and $\rho_{\rm crit}$, the critical density today. At large scales and in projection over a long redshift baseline, the factors $(k_x^2 - k_y^2)/k^2$ can be ignored. 

\subsection{Observables: Angular Power Spectra}
\label{sec:cells}

With Eq. \ref{eq:ia}, we are able to calculate the observables - the angular power spectra ($C_l$'s). We evaluate these power spectra with the latest version (v2.1.0) of the {\tt CCL} library \citep{pyccl}. Generically, the angular power spectrum between two tracers $a$ and $b$ of the large-scale structure is given by
\begin{equation}\label{eqn: ciaia}
    C_{\ell}^{a b}=\frac{2}{\pi} \int_{0}^{\infty} k^2 d k P_\phi(k,z) \Delta_{\ell}^{a}(k) \Delta_{\ell}^{b}(k),
\end{equation}
where ${ab \in \{n-IA, IA-IA, n-WL,}\}$, i.e. the relevant combinations of alignment shapes, galaxy number counts and lensing shapes for our work. $P_\phi$ is the power spectrum of the gravitational potential as a function of redshift. The $\Delta$'s are the relevant kernels for the observables we are correlating. For the number density ($n$) and alignment ($IA$), these are \citep{Hirata_2004,Schmidt_2012, Schmidt_2015}:
\begin{align}
\Delta_{\ell}^{n}(k)&=\int d z \frac{dN}{dz} b_n(z) T_\delta(k) j_{\ell}(k \chi(z))\label{eqn: ntracer}, \\\
\Delta_{\ell}^{IA}(k)&=-\sqrt{\frac{(\ell+2) !}{(\ell-2) !}} \int d z \frac{dN}{dz} A_{IA}(z) \frac{C_1 \rho_{\rm crit} \Omega_M }{D(z)} T_\delta(k) \frac{j_{\ell}(k \chi(z))}{[k \chi(z)]^{2}}\label{eqn:IAtracer},
\end{align}
where $b_n(z)$ is the galaxy bias; $j_l$, the spherical Bessel function; $T_\delta$, the transfer function, relating the potential power spectrum to the matter power spectrum; $\chi$, the comoving distance; and $dN/dz$, the redshift distribution of galaxies or clusters, normalized to integrate to $1$. To model the transfer function, we use the non linear power spectrum from CAMB \citep{camb}. For our purposes, this approximation is sufficient, as we mostly focus on large-scale alignments between halos. At small scales ($\lesssim 1-10 \operatorname{Mpc/h}$) and for galaxies, such approximation breaks down \citep{Singh_2015}, as galaxy separations become comparable to the halo size \citep{Schneider_2010,fortuna2020halo}.

We can also calculate these power spectra for the shape component due to weak gravitational lensing. The corresponding $\Delta_{\ell}^{WL}(k)$ for lensing is \citep{Schmidt_2012}:
\begin{equation}
\Delta_{\ell}^{WL}(k)=-\frac{1}{2} \sqrt{\frac{(\ell+2) !}{(\ell-2) !}} \int \frac{d z}{H(z)} W^{WL}(z) T_{\phi+\psi}(k, z) j_{\ell}(k \chi(z)),
\end{equation}
where $W^{WL}$ is the weak lensing kernel, given by
\begin{equation}
    W^{WL}(z) \equiv \int_{z}^{\infty} d z^{\prime} \frac{dN}{dz'}\left(z^{\prime}\right) \frac{\chi^{\prime}-\chi}{\chi^{\prime} \chi}.
\end{equation}
In standard $\Lambda$CDM, $T_{\phi+\psi}$ is related to the matter transfer function by 
\begin{equation}
    T_{\delta}=-\frac{k^{2}}{3 H_{0}^{2} \Omega_{m}} \frac{T_{\phi+\psi}}{1+z}.
\end{equation}

\subsection{Fisher forecast \& Signal-to-Noise}
\label{sec:errors}

Our aim is to predict and compare the overall detectability of alignment correlations of clusters and galaxies in several set ups. To this end, we perform a Fisher forecast for the signal-to-noise $S/N$ of the predicted angular power spectra between two probes $a$ and $b$, $C^{a b}_l$. The signal-to-noise is given by:
\begin{equation}\label{eqn:sn}
\frac{S}{N} = \sqrt{\sum_{l=l_{\min}}^{l=l_{\max}}\frac{\left(C^{a b}_l\right)^2}{\operatorname{Var}[C^{a b}_l]}}.
\end{equation}
Here, $\operatorname{Var}[C^{a b}_l]$ is the variance of the power spectrum. We assume that all the perturbations are Gaussian, then the variance is given by \citep{cov}: 
\begin{equation}\label{eqn:cov}
    \operatorname{Var}[C^{a b}_l] = \frac{\tilde{C}^{a a}_l\tilde{C}^{b b}_l + 
    \tilde{C}^{a b}_l\tilde{C}^{b a}_l}{(2l+1)f_{\operatorname{sky}}}.
\end{equation}
The factor $f_{sky}$ is the fraction of the sky the survey has observed. $\tilde{C}^{a b}_l$ is the power spectrum including noise. We consider noise for the auto-power spectra, and we assume the noise terms to be white, which is warranted under the assumption of a uniform survey \citep{Troxel18}. This results in the following power spectra for shape and number density \citep{Duncan_2013}: 
\begin{equation}\label{eqn:noise}
    \tilde{C}^{\gamma \gamma} = C^{\gamma\gamma}_l +  \frac{\sigma_\gamma^2}{N_\gamma} \ \ \ \ \ \ \ \tilde{C}_l ^{nn} = C_l^{nn} + \frac{1}{N}
\end{equation}
Here, $\sigma_\gamma$ is the shape dispersion per component, $N$ is the amount of galaxies/clusters per steradian, $N_\gamma$ is the number of galaxies or clusters with accurate shape measurements. In practice, we take $N$ and $N_\gamma$ to be the same in this work. We can thus also make predictions for the fractional error, $\delta C^{a b}_l/C^{a b}_l=\sqrt{{\rm Var}[C^{a b}_l]}/C^{a b}_l$. For $l_{\min}$ and $l_{\max}$, we use $2 \leq l \leq 1000$ in line with our modelling assumptions. This is a conservative range, considering that current and future surveys can access a larger range of scales \citep{Krause_2015}. 

\subsection{Galaxy and Cluster samples}
\label{sec:distributions}

For our predictions of the noise (Eq. \ref{eqn:noise}) and the $C_l$'s (Eq. \ref{eqn: ciaia}), we need estimates of $A_{IA}$, $b_n$, $dN/dz$, $N$ and $\sigma_\gamma$. We obtain these from the current SDSS survey \citep{SDSS}, and forecast them for the opcoming LSST survey.

\subsubsection{SDSS cluster sample}
\label{sec:sdssclusters}

For SDSS, we use the publicly available {\tt redMaPPer} catalogue \citep{Rykoff_2014}. There are 26111 clusters in the range $0.08<z\leq 0.6$ and over the richness range $20 \leq \lambda \leq 200$. Where the richness $\lambda$ is a probabilistic estimate of the number of galaxies in the cluster. $\lambda$ is correlated with luminosity, thus this also includes a luminosity cut. The shape is obtained via the quadruple moment $Q_{ij}$ following \citep{Vanuitert_2017}, this is calculated by looking at the locations of galaxies with respect to the Brightest Cluster Galaxy (BCG) as follows 
\begin{equation}
    Q_{ij} = \frac{\sum_k(\theta_{i,k} - \theta^{BCG}_i)(\theta_{j,k} - \theta^{BCG}_j)p_{mem,k}}{\sum_k p_{mem,k}}, \ \ i,j \in \{1,2\}.
    \label{eqn: transform}
\end{equation}
Here, $\theta$ is the angular coordinate and $p_{mem}$ the probability that the galaxy is member of the cluster. With the quadruple moment at hand, a complex ellipticity $\epsilon$ can be calculated as
\begin{equation}
    \epsilon= \epsilon_1+i \epsilon_2=\frac{Q_{11}-Q_{22}+2 \mathrm{i} Q_{12}}{Q_{11}+Q_{22}+2 \sqrt{Q_{11} Q_{22}-Q_{12}^{2}}}.
\end{equation}
The components $\epsilon_1$ and $\epsilon_2$ can then be obtained.

We thus also obtain the standard deviation of the ellipticities, $\sigma_\gamma$, directly from the {\tt redMaPPer} catalogue. For the full richness and redshift range this is $\sigma_\gamma=0.098$, using $\sigma_\gamma^2 = \frac{1}{2}(\sigma_{\epsilon_1}^2 + \sigma_{\epsilon_2}^2)$. When binning the data by richness, we re-calculate this accordingly, though this procedure does not have a significant impact on our results.

\begin{figure}
    \centering
    \includegraphics[width = .4\textwidth]{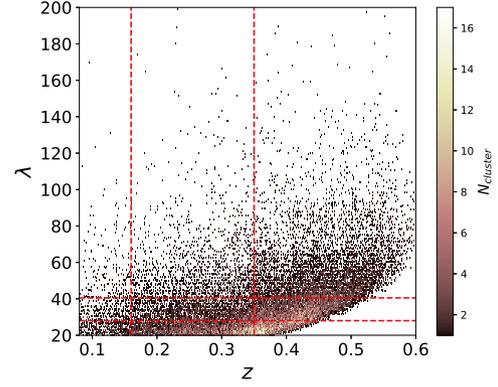}
    \caption{The SDSS {\tt redMaPPer} distribution seen as a two-dimensional histogram of richness, $\lambda$, and redshift, $z$. The color scale indicates the number of clusters. The dashed lines delineate the bins in which the alignment signal was measured by \citet{Vanuitert_2017}. The sample is mostly complete at $z\lesssim 0.35$. We rely on this sample to estimate the $S/N$ in cluster alignments currently available.}
    \label{fig:2dhist}
\end{figure}

The distribution of the number counts as a function of richness $\lambda$ and redshift $z$ are obtained directly from the catalogue and shown in Figure \ref{fig:2dhist}. Below $z \approx 0.35$, the data is mostly complete with the increasing number counts resulting as a consequence of the increasing comoving volume at higher redshift. Above this redshift, we see the effects of a selection bias: lower richness clusters are harder to detect, as mentioned by \citet[sec. 10 and 11]{Rykoff_2014}. 

For $b_n$ and $A_{IA}$, we rely on the measured values from \citet{Vanuitert_2017}. They measured an alignment amplitude of $A_{IA}^{gen} = 12.6^{+1.5}_{-1.2}$ at a pivot redshift of $z_0=0.3$, with a richness $\lambda$ and redshift $z$ dependence of the amplitude parameterized as:
\begin{equation}\label{eqn:clamp}
A_{IA}(\lambda,z) = A_{IA}^{gen} \left(\frac{1+z}{1+z_0}\right)^\eta \left(\frac{\lambda}{\lambda_0}\right)^\beta,
\end{equation}
where $\lambda_0$ is a pivot richness of $30$. For $\eta$ and $\beta$, their fitted values were: $\eta = -3.2^{+1.31}_{-1.40}$ and $\beta = 0.6^{+0.20}_{-0.27}$.
For $b_n$, we use the value they measured in the redshift cut $0.16 < z \leq 0.35$, this is $b_n = 4.25^{+0.15}_{-0.16}$. This is a reasonable generalization, as the measured values in the other cuts are within those error bars. All the measured values can be found in Table 1 from \citet{Vanuitert_2017}. 

\subsubsection{LSST cluster sample}
\label{sec:lsstclusters}

The expected number counts as a function of redshift and solid angle for LSST is given by \citep{Eifler}
\begin{equation}
    \frac{dN}{dz d\Omega} = \frac{dV_c}{dz d\Omega} \int dM\frac{dn}{dM}\int_{\ln\lambda_{\min}}^{\ln\lambda_{\max}}d\ln\lambda \ p(\ln\lambda|M,z),
\end{equation}
where $\frac{dn}{dM}$ is the halo mass function from \cite{tinker}, giving the number of dark matter halos per unit mass and comoving volume at a certain redshift. $\frac{dV_c}{dz d\Omega}$ is the comoving volume element and $p(\ln\lambda|M,z)$ is a lognormal distribution relating the observed richness to underlying mass, given by
\begin{equation}
     p(\ln \lambda | M, z)=\frac{1}{\sqrt{2 \pi} \sigma_{\ln \lambda | M, z}} \exp \left[-\frac{(\ln \lambda-\langle\ln \lambda\rangle(M))^{2}}{2 \sigma_{\ln \lambda | M, z}^{2}}\right].
\end{equation}
This distribution is motivated by the weak lensing mass calibration of SDSS clusters \citep{murataa}. The same work constrained the mass-richness relation to be
\begin{equation}\label{eqn:massrichness}
    \langle\ln \lambda\rangle(M, z | A, B, C)=A+B \ln \left(\frac{M}{M_{piv}}\right)+C \ln (1+z),
\end{equation}
where the pivot mass is defined as $M_{piv}= 3\times10^{14}M_{\odot}/h$, $A= 3.207 \pm 0.045$, $B=0.993 \pm 0.045$ and $C=0.0 \pm 0.3 $. We follow the approach of the DESC science requirements document \citep{DESC} and use the scatter from \cite{murataa} with $q_m$ set to 0, i.e.
\begin{equation}
    \sigma_{\ln \lambda | M}\left(M, z | \sigma_{0}, q_{z}\right)=\sigma_{0}
    +q_{z} \ln (1+z).
\end{equation}
Here, $\sigma_0=0.456 \pm 0.045$ and $q_z=0 \pm 0.1$. Notice that our relation between richness and mass does not depend on redshift. For LSST, we use the same richness range as for SDSS, namely $20 \leq \lambda \leq 200$. 

The total number counts are given by the integral over redshift, multiplied by the sky area covered, $\Omega=4\pi f_{\rm sky}$,
\begin{equation}\label{eq:nclu}
N = \Omega \int_{z_{min}}^{z_{max}} dz\frac{dN}{dz d\Omega}.
\end{equation}
Eq. \ref{eq:nclu} does not account for selection effects, i.e. the fact that it becomes more difficult to detect lower $\lambda$ clusters at higher redshifts. This effect could have several possible origins: a Malmquist bias or the paucity of red cluster members, for example \citep{Rykoff_2014}. As discussed previously, the SDSS {\tt redMaPPer} clusters clearly show such a selection bias (Figure \ref{fig:2dhist}). 

\begin{figure}
	\centering
	\subfloat[Cluster redshift distribution]{\includegraphics[width=0.35\textwidth]{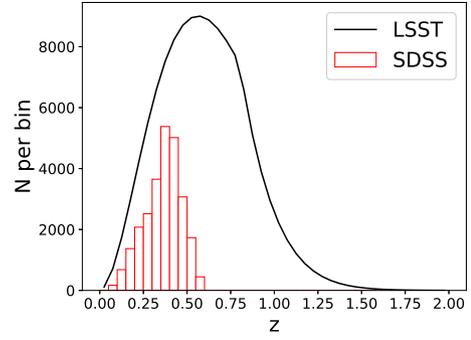}\label{fig:histlsst}}
	\qquad
	\subfloat[Galaxy redshift distribution LSST]{\includegraphics[width=0.35\textwidth]{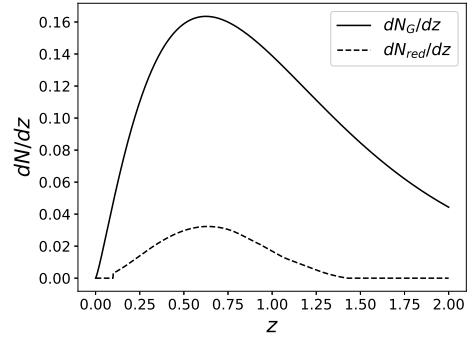}\label{fig:galaxieslsst}}
    \caption{Panel (a) shows the number of galaxy clusters as a function of redshift for LSST (solid black), compared to the {\tt redMaPPer} SDSS sample distribution (red histogram). Note that the break at $z = 0.8$ is due to the selection effects we include. The $y$ axis represents the number of clusters per 0.05 $z$-bin. Panel (b) shows the galaxy number count distribution with redshift for LSST (solid line), and the distribution of red galaxies among them (dashed, from \citealt{Schmidt_2015}). The normalisation is arbitrary.}
    \label{fig:distributions}
\end{figure}

To model a selection effect in the case of LSST, we use a lower richness limit that increases with redshift, i.e. 
\begin{equation}\label{eq:seleff}
    \lambda_{\min}(z) = \lambda_{\min}^0 + a(z-z_0)\Theta(z-z_0).
\end{equation}
In Eq. \ref{eq:seleff}, $\lambda_{\min}^0$ is the lower richness limit without selection effects, $a$ regulates a linear dependence of the richness limit with redshift, $\Theta(z-z_0)$ is the Heaviside step-function, and $z_0$ is the redshift above which the selection effect becomes important. We conservatively estimate $z_0 = 0.8$ for LSST and  $a = 40$ based on extrapolation of the {\tt redMaPPer} data from SDSS \citep{Rykoff_2014, Vanuitert_2017} and the preliminary Year 1 data from the Dark Energy Survey \citep{desart,des1,redmlensing}. 
We obtain a number count of $\simeq 1.2 \times 10^5$ clusters for a redshift range $0\leq z \leq 2$.  The redshift distribution is shown in Figure \ref{fig:histlsst}.

\begin{figure}
     \centering
     \includegraphics[width = 0.4\textwidth]{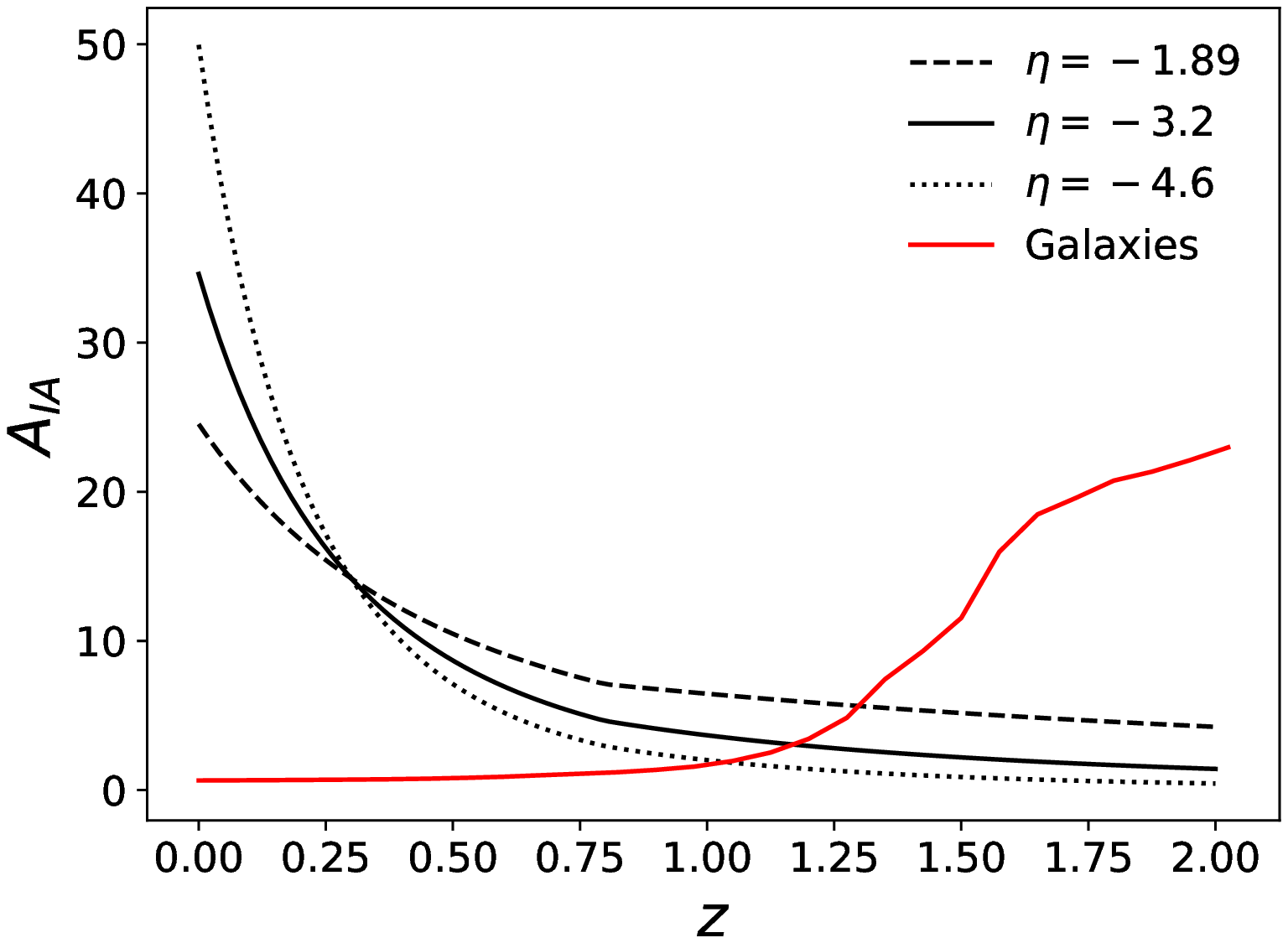}
     \caption{The alignment amplitude of both the LSST clusters (extrapolated from \citealt{van_Uitert_2018}) and galaxies (as predicted in \citealt{Schmidt_2015}). The alignment amplitude of the clusters is also shown for two values of $\eta$ at 1$\sigma$ distances from the best-fit $\eta = -3.2$. The steep increase in the alignment amplitude of galaxies is caused by a Malmquist bias, i.e. the fact that more luminous galaxies are observed at high redshift when the survey is apparent magnitude-limited.}
     \label{fig:aia}
\end{figure}

The cluster bias is obtained by multiplying $p(\ln\lambda|M,z)$ with the halo bias function $b_h$ and then normalising the result as follows \citep{Eifler}: 
\begin{equation}
    b_n(z)=\frac{\int \mathrm{d} M \frac{\mathrm{d} n}{\mathrm{d} M} b_{\mathrm{h}}(M)\int_{\ln \lambda_{ \min }}^{\ln \lambda_{\max }} \mathrm{d} \ln \lambda p(\ln \lambda | M)}{\int \mathrm{d} M \frac{\mathrm{d} n}{\mathrm{d} M} \int_{\ln \lambda_{ \min }}^{\ln \lambda_{\max }} \mathrm{d} \ln \lambda p(\ln \lambda | M)}
    \end{equation}
We use the halo bias function from \cite{tinker}.

The mean richness per redshift is then 
\begin{equation}\label{eqn:lambda}
\lambda(z)=\frac{\int \mathrm{d} M \frac{\mathrm{d} n}{\mathrm{d} M} \int_{\ln\lambda_{min}}^{\ln\lambda_{max}}d\ln\lambda \ \lambda p(\ln\lambda|M,z)}{\int \mathrm{d} M \frac{\mathrm{d} n}{\mathrm{d} M} \int_{\ln \lambda_{ \min }}^{\ln \lambda_{\max }} \mathrm{d} \ln \lambda p(\ln \lambda | M)}
\end{equation}
All the integrals over $p(\ln \lambda | M)$ are performed analytically; the integral over the halo mass function is then done numerically. 

We use the constraints on alignment amplitude from \citet{Vanuitert_2017}, given by Eq. \ref{eqn:clamp}. The prediction is shown in Figure \ref{fig:aia}. The parameter whose uncertainty has the most impact on the predictions is the slope of the redshift evolution power-law. We show the impact of varying it within its $68\%$ confidence level in Figure \ref{fig:aia}.

For LSST we adopt the same shape noise as for the SDSS clusters, namely $\sigma_\gamma = 0.098$ clusters in the richness range $20 \leq \lambda \leq 200$. In this case, this is not varied with redshift. 

\subsubsection{SDSS galaxy sample}
\label{sec:sdssgalaxies}

We are interested in comparing the detectability of cluster alignment with predictions for the corresponding galaxy alignment statistics. Alignments are significantly measured in several samples nowadays, with some of the most stringent constraints on the alignment model given by the SDSS-III BOSS LOWZ sample \citep{Singh_2015}. To mimic the redshift distribution of this sample, we assume the number of galaxies per comoving volume to be constant:
\begin{equation}
    \frac{dN}{dz} \propto \frac{dV_c}{dz d\Omega} = D_H\frac{(1+z)^2 D_A(z)^2}{\sqrt{\Omega_M (1+z)^3 + \Omega_k (1 + z)^2 + \Omega_\Lambda}},
\end{equation}
with $D_H=c/H_0$ the Hubble distance, and $D_A(z)$ the angular diameter distance. This distribution corresponds to a volume-limited survey and is in line with the properties of the LOWZ sample. In line with the LOWZ sample, we adopt a redshift range spanning $0.16 \leq z \leq 0.36$.

We use the values for $b_n$, $A_{IA}$ and $\sigma_\gamma$ from \cite{Singh_2015}: $b_n = 1.77 \pm 0.04$, $A_{IA} = 4.6 \pm 0.5$ and $\sigma_\gamma=0.2$.

\subsubsection{LSST red galaxy sample}
\label{sec:lsstgalaxies}

To compare to our LSST cluster sample, we also need a galaxy sample for LSST. We use the results of \citet{lsstdist} for the distribution of the total galaxy number count, 
\begin{equation}\frac{dN}{dz}\propto z^{\alpha} \exp \left[-\left(\frac{z}{z_{0}}\right)^{\beta}\right].\end{equation}
Here, $\alpha$ = 1.21, $\beta = 1.05$ and $z_0 = 0.5$. The distribution has a median redshift $z_m = 0.82$. 

For alignments, we are however only interested in the red fraction of galaxies. We rely on earlier work \citep{Schmidt_2015,shapes} to model the alignment amplitude and the red fraction of the galaxy population. The corresponding redshift distribution is shown in Figure \ref{fig:galaxieslsst}. The predicted red fraction suggests $2.6$ red galaxies per arcmin$^2$ will be observed by LSST. We approximate the galaxy bias as \citep{DESC}:
\begin{equation}
    b_n(z) = \frac{0.95}{D(z)}.
\end{equation}

The alignment amplitude is modelled by using the correlation with luminosity found \citep{Joachimi_2011}, taking into account the fact that a magnitude-limited survey observes more luminous galaxies at higher redshifts, i.e. Malmquist bias. The correlation with luminosity gives rise to a very steep increase in the alignment amplitude with redshift, as shown in Figure \ref{fig:aia}. As in \citet{shapes}, we assume a shape dispersion $\sigma_{\gamma} = 0.26$. 
\section{Results}
\label{sec:results}

\begin{figure}
    \centering
    \includegraphics[width = 0.48\textwidth]{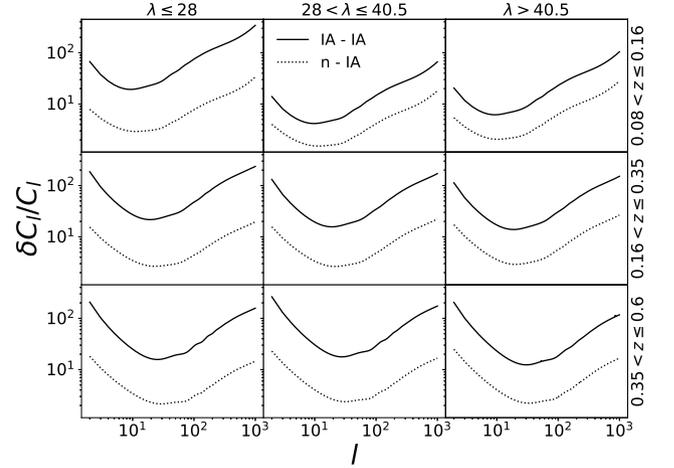}
    \caption{Fractional uncertainties in the alignment angular spectra of the SDSS clusters as a function of angular multipole. We adopted the same redshift and richness binning and the best-fit alignment model from \citet{van_Uitert_2018}. The solid line shows the alignment auto-correlation, while the dashed line corresponds to the cross-correlation of cluster positions and shapes.}
    \label{fig:fracersuitert}
\end{figure}

We are interested in the trade-off between galaxy and cluster alignments in currently available samples, such as in SDSS. We adopt several cuts in richness and redshift for the {\tt redMaPPer} sample, as shown by the red dashed lines in Figure \ref{fig:2dhist}, to see how the predicted fractional errors on the alignment angular power spectra vary. The results are shown in Figure \ref{fig:fracersuitert}, each panel corresponding to a given selection in redshift and richness. Both $n-IA$ and $IA-IA$ fractional uncertainties are presented in each panel. As expected, uncertainties in $n-IA$ are always lower than for $IA-IA$ given the impact of shape noise.

The fractional errors on the cluster alignment power spectra increase with redshift, as can be seen by comparing the rows in Figure \ref{fig:fracersuitert}. They also do not vary significantly with richness, as can be seen by comparing the columns of Figure \ref{fig:fracersuitert}. The only major difference is in the first redshift bin, however we note that in this bin the uncertainties on the alignment constraints by \citet{Vanuitert_2017} are significant. The fractional error being roughly constant over the richness bins suggests that the enhanced alignment amplitude is offset by the reduced number counts in the higher richness bins. 

\begin{figure}
	\centering
	\subfloat[]{\includegraphics[width=0.48\textwidth]{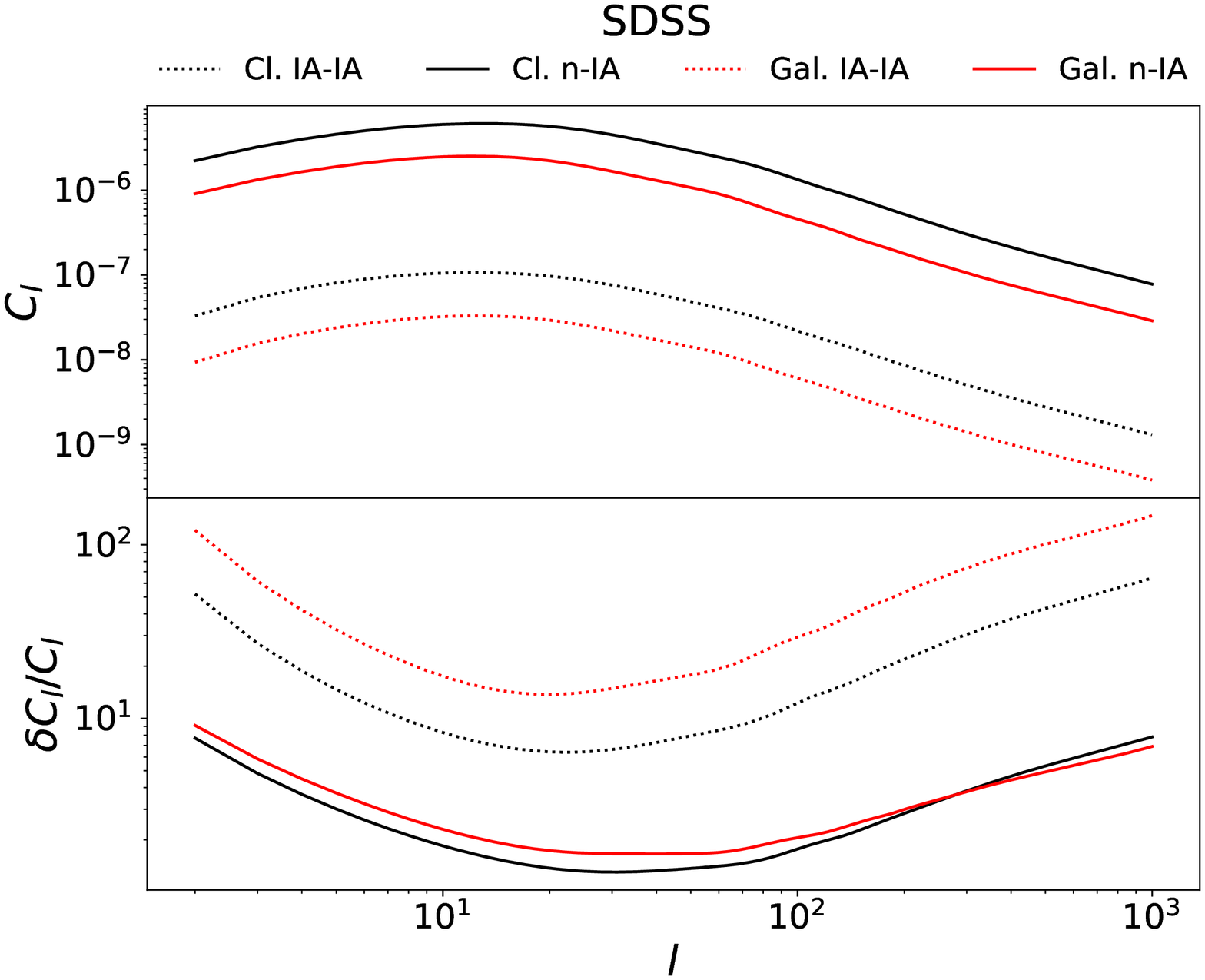}\label{fig:cellssdss}}
	\qquad
	\subfloat[]{\includegraphics[width=0.48\textwidth]{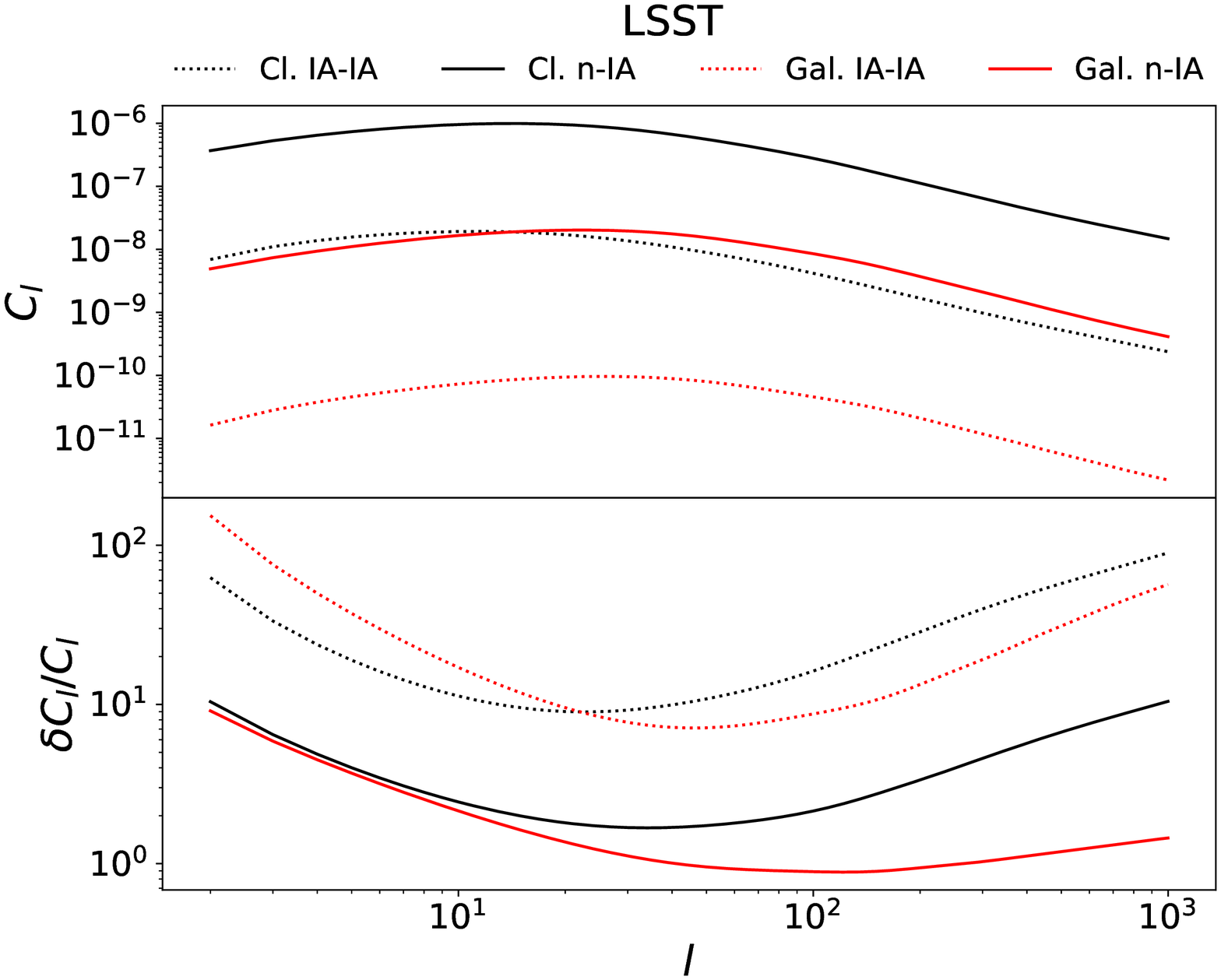}\label{fig:cellslsst}}
    \caption{Panel (a) shows the angular power spectra (top sub-panel) and their expected fractional uncertainties (bottom sub-panel) for the auto-correlation of intrinsic shapes ($IA-IA$, dotted line) and the cross-correlation of positions and shapes ($n-IA$, solid line) for the SDSS cluster (black) and galaxy samples (red). (Notice that we show the absolute value for $n-IA$, since the sign of the alignment signal does not impact our $S/N$ estimations.) Panel (b) shows analogous results for LSST.}
\end{figure}

In Figure \ref{fig:cellssdss}, we compare the alignments of galaxies and clusters in SDSS. The computation of the expected uncertainties, presented in the lower panel, allows us calculate the S/N according to Eq. \ref{eqn:sn}:
\begin{itemize}
    \item For the auto-correlation of intrinsic shapes, $IA-IA$ (dashed black), we see that clusters present an advantage over the galaxies (dashed red): the signal-to-noise for the clusters is $1.5$, in comparison to $0.66$ with galaxies. 
    \item  For the cross-correlation of positions and shapes, $n-IA$, the fractional errors are comparable between clusters (solid black) and galaxies (solid red). We find that clusters have a $S/N$ of $9.7$ while galaxies have the slightly lower value of $8.9$. 
\end{itemize}
The reason why the cluster $IA-IA$ is especially significant in comparison to galaxies is that the signal-to-noise (Eq. \ref{eqn:sn}) depends on $A_{IA}$ (Eqs. \ref{eqn: ciaia} and \ref{eqn:IAtracer}) to quadratic order, while for $n-IA$ it does so at linear order. Clusters also have a higher $A_{IA}$ and lower $\sigma_\gamma$ than galaxies, and this is particularly impactful in the $IA-IA$ signal. 

We repeat this procedure for LSST to find out where this trade-off between cluster and galaxy alignments lies in a more futuristic survey. The signal and the fractional errors are shown in Figure \ref{fig:cellslsst}. In this case, and compared to SDSS, clusters no longer present a significant advantage as an alignment tracer over the galaxies. 
\begin{itemize}
    \item For the $IA-IA$ signal (dashed black), the clusters are still comparable in fractional uncertainties with the galaxies (dashed red); the latter providing a slightly better result. Clusters have a signal-to-noise of $1.1$ compared to $1.8$ for galaxies.
    \item For the $n-IA$ signal the fractional errors of the galaxies are significantly better (solid red vs. solid black). We find that clusters have a $S/N$ of $7.9$, which is considerably lower than the value of $28$ we find for galaxies.
\end{itemize}

\begin{figure}
	\centering
	\subfloat[]{\includegraphics[width=0.48\textwidth]{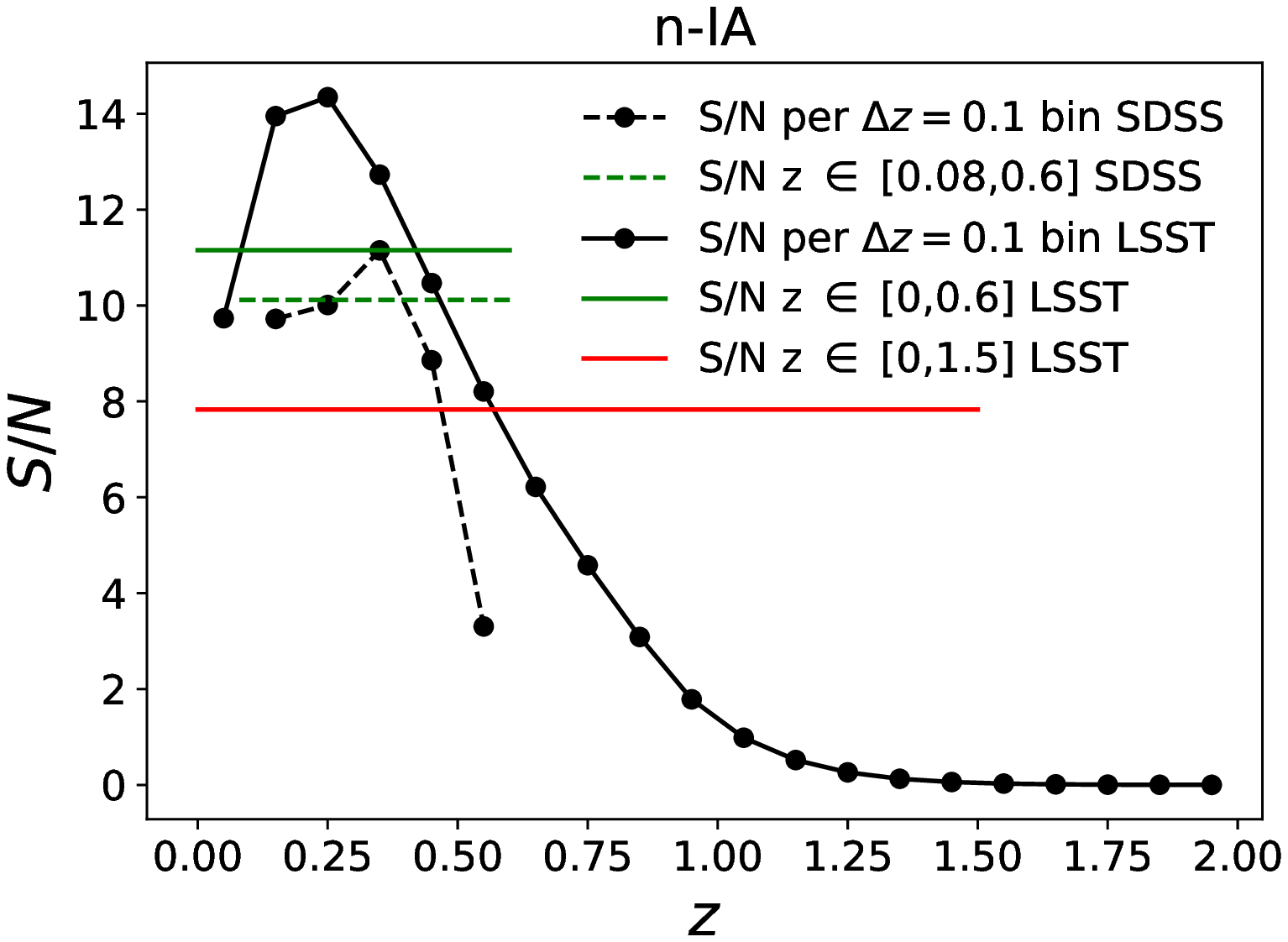}\label{fig:sn1}}
	\qquad
	\subfloat[]{\includegraphics[width=0.48\textwidth]{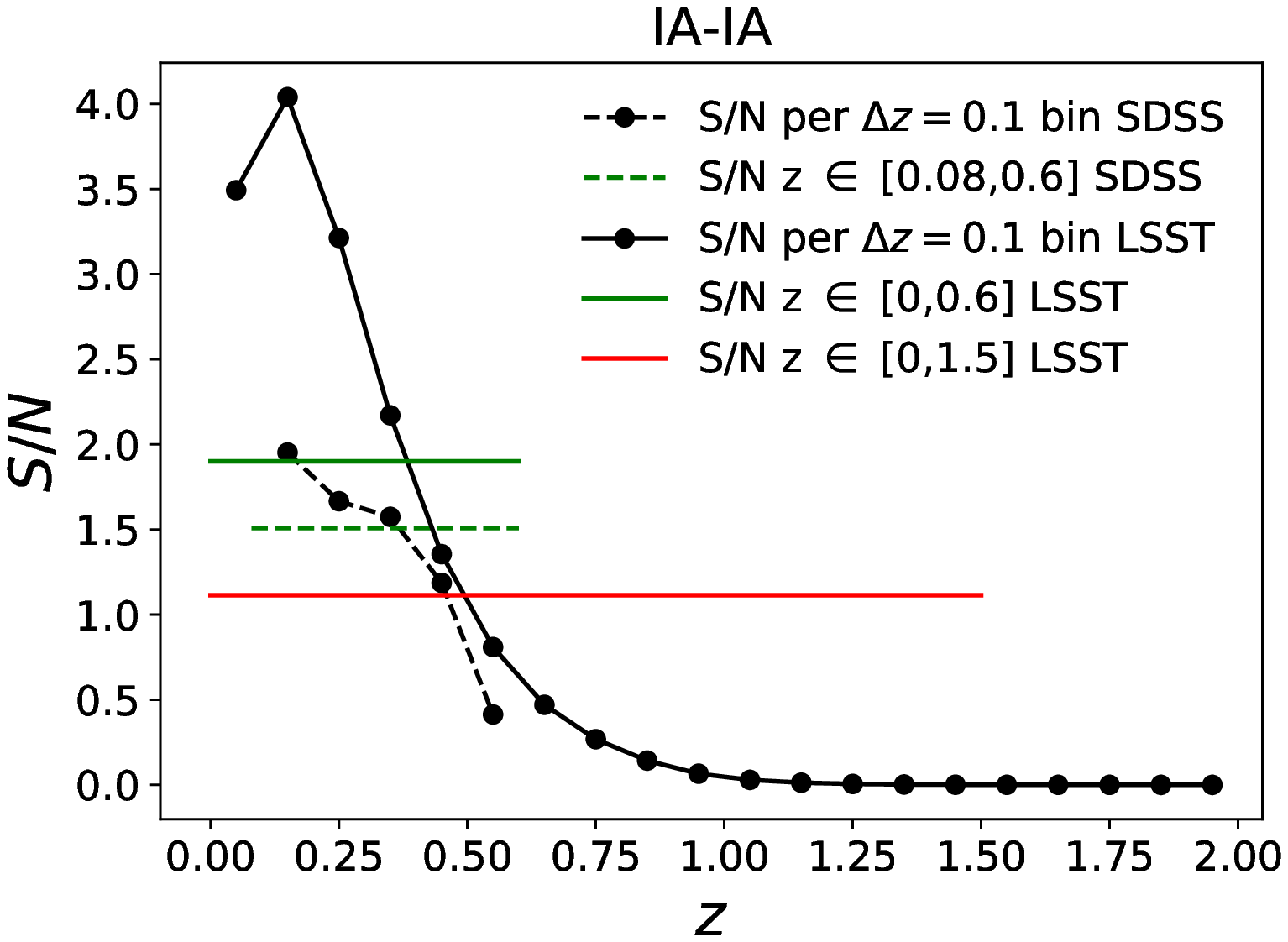}\label{fig:sn2}}
    \caption{The signal-to-noise ($S/N$) for the $n-IA$ (panel a) and $IA-IA$ (panel b) angular power spectra for the LSST/SDSS clusters calculated in various redshift bins. The black line shows the $S/N$ calculated for evenly spaced redshift bins with a width of $\Delta z = 0.1$ for LSST. The red line gives the value of the $S/N$ when calculated for one bin with the redshift range $[0,1.5]$ for LSST; the green line corresponds to the $S/N$ in the range $[0,0.6]$ for LSST. The dashed black line shows the S/N in $\Delta z = 0.1$ bins for SDSS, where we skip the first bin as {\tt redMaPPer} has a lower bound of $z=0.08$. The dashed green line corresponds to the $S/N$ in the range full redshift range ($z=[0.08,0.6]$) of SDSS data.}
    \label{fig:sn}
\end{figure}

An interesting feature of the results is that for both correlations in the LSST cluster sample, the signal-to-noise is actually smaller than that of their counterpart in the SDSS cluster sample. At first sight this is peculiar, as LSST is expected to observe about 5 times as many clusters and the sample should in fact include the SDSS clusters (see Figure \ref{fig:histlsst}). To investigate this further, we have calculated the signal-to-noise in several redshift bins for  both SDSS and LSST. This is shown in Figure \ref{fig:sn}.

Focusing first on the LSST results shown in Figure \ref{fig:sn}, we have chosen 3 binning schemes: the first bin is a single bin that spans the range $z < 1.5$ (red solid curve); the second bin is also a single bin comparable with the SDSS redshift range going from $z=0$ to $z=0.6$  (green solid curve); third, we have calculated the signal-to-noise in 20 bins with width $\Delta z = 0.1$ to localize where the signal-to-noise peaks (black solid curve). Isolating clusters closer to each other (black solid curve) increases the signal strength, but it also reduces the number counts in a given bin. 

From Figure \ref{fig:sn}, we also see that the signal-to-noise of clusters is mostly located at lower redshift (at $z\lesssim 0.6$). When calculating the angular power spectra, the clusters at higher redshift do not contribute significantly to the signal. Rather, they tend to wash it out when brought into the measurement. If we focus on the $[0,0.6]$ bin, we observe that the $S/N$ is better than SDSS in this bin (green dashed curve), as expected. Here, the $IA-IA$ signal-to-noise is actually better than the LSST signal-to-noise for the full redshift range. Most of the signal seems to be located around $z=0.3$.

For the fiducial results presented above, we have considered the best-fit value for $\eta$ as measured by \citet{Vanuitert_2017}. Varying this parameter within its $1\sigma$ uncertainty, as is shown in Figure \ref{fig:aia}, can yield a slight increase in $S/N$ of $0.2$ for the $IA-IA$ correlation, and of $1.3$ for the $n-IA$ correlation when calculated in the complete redshift range. This is however insufficient to drive it higher than the galaxy alignment $S/N$ in the same redshift range.

\section{Weak Lensing of Galaxy Clusters}
\label{lensing}

The effects of weak lensing and intrinsic alignments are always measured jointly in projection, as indicated by our Eq. \ref{eq:shapesum}. Both phenomena contribute to distorting the shape of a cosmological object and need to be modelled jointly. An exception is the very low redshift regime, where lensing tends to be subdominant compared to intrinsic alignments, due to the lack of intervening structures in the path of the source photons.
For these reasons, it is important to assess at what level can cluster shapes be lensed. The shape of the cluster, as measured by the estimator \ref{eqn: transform}, is subject to lensing distortions due to the coherent angular displacement of the member galaxies, i.e. the magnification effect. 

\begin{figure}
    \centering
    \includegraphics[width=0.5\textwidth]{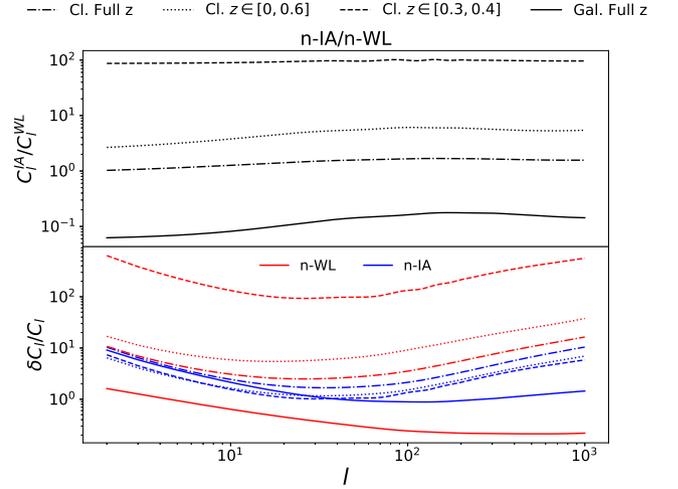}
    \caption{Ratio of alignment to weak lensing angular position-shape power spectra for both galaxies and clusters in LSST (top panel), and the fractional uncertainties predicted for both signals (bottom panel). The solid curve indicates predictions for galaxies in the LSST survey. Other curves indicate predictions for cluster observables. From the dot-dashed to the the dotted and the dashed line, we place more restrictive redshift cuts. These progressively allow for better isolation of the alignment signal. (Notice that we show the absolute value for $n-IA$, since the sign of the alignment signal does not impact our $S/N$ estimations.) Overall, we find that cluster alignments present a significant advantage over galaxy alignments for cosmological applications that rely on this observable, due to the lower contribution of the lensing component.} 
    \label{fig:clusterlensing}
\end{figure}

As we have seen in previous sections, alignments are stronger for clusters than for galaxies, thus we expect that the fractional contribution of gravitational lensing to the total shape correlations should be smaller. We estimate this contribution using the formalism of Section \ref{sec:cells} and show the results in Figure \ref{fig:clusterlensing} as the ratio of weak lensing angular power spectra to the corresponding alignment statistic. We restrict to the case of position-shape correlations for LSST clusters and draw a comparison with the red galaxy sample. (For SDSS, the alignment signal is usually stronger than the lensing signal for the samples we consider.) Shape-shape correlations are not only noisier, as seen in Figure \ref{fig:sn2}, but it also becomes harder to isolate the alignment signal due to the presence of alignment-lensing cross-correlations.

Figure \ref{fig:clusterlensing} shows that the fractional contribution of intrinsic alignment to total shape correlations is much larger in the case of clusters (by a factor from $\sim$10 to $\sim$1000). The lower panel also shows that the typical uncertainty in the alignment measurements (blue curves) is lower than in the weak lensing measurement (red curves) for this sample in LSST. This trend is reversed in the case of red galaxies in LSST. Moreover, the alignment signal is better constrained when narrow redshift ranges are chosen, in agreement with Figure \ref{fig:sn1}.

We conclude that cosmological applications of intrinsic alignments might benefit from using clusters instead of galaxies as tracers, since their alignment can be better isolated.

\section{Discussion}
\label{sec:discussion}

Our forecasts for cluster and galaxy alignments in SDSS and LSST suggest that the cluster alignment signal is particularly strong and competitive with galaxy alignment at low redshift. This is particularly true of the $IA-IA$ signal, although the overall $S/N$ remains low. Nevertheless, it might be possible to detect the cluster $IA-IA$ signal leveraging on the selection of close pairs, as suggested in Figure \ref{fig:sn}. \citet{Blazek_2011}, in fact, obtained a detection of the $IA-IA$ signal for galaxies by using a correlation function estimator that isolates galaxies over a restricted range along the line of sight \citep{wfunctie}. Figure \ref{fig:sn} suggests that these type of estimators would be advantageous in studies of intrinsic alignments, compared to typical angular power spectra.

At higher redshift the signal-to-noise of cluster alignments quickly fades away. This can be explained by the alignment amplitude $A_{IA}$ for clusters being a decreasing function with redshift \citep{Vanuitert_2017}, as opposed to the case for galaxies, and even when selection effects on the cluster sample are considered (Figure \ref{fig:aia}). 
The difference in the redshift dependence for galaxies and clusters is quite remarkable, especially at the redshift where their alignment amplitudes become comparable. \citet{Vanuitert_2017} found quite a steep redshift dependence $\eta$ for clusters. For galaxies there also seem to be hints of a negative redshift dependence, however this is not as clearly observed \citep{Samuroff_2019}. 
To render this type of predictions more accurate, it is important that the factor $\eta$ is well constrained for both galaxies and clusters. Connecting the observed galaxy and cluster alignment measurements to the results from simulations would also shed light on the origin of their redshift dependence. 

In our work, we have relied on the fiducial alignment amplitude measured by \citet{Vanuitert_2017}. These authors explored the impact of a number of systematic effects on cluster alignment measurements. Miscentering of clusters \citep{Rozo14} and cluster membership selection \citep{Zu17} were deemed to only play a role at small separations. Thus, we do not expect these to significantly impact our predictions. On the other hand, \citet{Vanuitert_2017} highlighted that typical photometric redshift errors on the cluster redshift can drive a $0.5-1\sigma$ suppression of the alignment signal at our scales of interest. We hypothesize this could also have an impact on redshift dependence for the alignments of the clusters. In general, all of these systematic effects could be explored further using dedicated simulations for future surveys. This would allow for a prediction of the $A_{IA}$ parameter that is tailored to a given survey and for a more realistic assessment of the noise contribution, including the impact of survey geometry and potential non-Gaussian terms in the covariance \citep{Takada1,Takada2}.

In the future galaxy cluster alignment could be useful for several applications.
\begin{itemize}
    \item At low redshift, clusters have a very favourable signal-to-noise, this could potentially be useful for the extraction of cosmological information. 
    \item Clusters can help improve our understanding of the alignment mechanism in general, and in particular at higher mass scales than galaxies. 
    \item The shape of a galaxy cluster traces the shape of the dark matter halo \citep{haloshape1, haloshape2}. Cluster alignments could thus provide a proxy for studying the alignment of the dark matter. This would be particularly relevant in the context of galaxy-halo alignment models. Recent numerical simulations suggest that galaxies gain their alignment by ``catching up'' with their halos over time \citep{Chisari_2017,Bhowmick_2019} but this trend remains unconstrained from observations. 
    \item At high redshift, correlations between galaxy shapes are dominated by the weak lensing effect. This makes it hard to isolate galaxy alignments. The shapes of clusters would be relatively less affected by this, with the result that it should be easier to extract the alignment part of the shape (Figure \ref{fig:clusterlensing}). This could help our understanding of intrinsic alignments in general, but it also makes it potentially easier to extract cosmological information from the signal. Our predictions on the contribution of magnification to cluster shape correlations could be tested in lightcone simulations.
\end{itemize}

In this work, we have explored cluster shape auto-correlations and the cross-correlation between cluster positions and shapes. We have not explored the cross-correlations between galaxies and clusters, which could add information in an scenario in which we would like to extract cosmological information from these phenomena. Another potential extension of this work is to study the benefits of a second shape estimator for clusters. Galaxy outskirts are typically more aligned than their inner regions \citep{Singh2}, with potential benefits to cosmological signal extraction \citep{shapes}. If this is a feature of tidal alignments in general, we expect a second measurement of cluster shapes could improve cosmological signal extraction. 

\section{Conclusions}
\label{sec:conclusion}

We have presented predictions for alignment of galaxies and clusters for the Sloan Digital Sky Survey (SDSS) and the Legacy Survey of Space and Time (LSST) of the Vera Rubin Observatory. We find that clusters have a very good signal-to-noise in intrinsic alignments at low redshift in comparison with galaxies. High redshift clusters however wash out the signal. This is partially solved by isolating clusters closer together in narrow redshift bins. Adopting the right selection can significantly improve the signal and make cluster alignment more competitive with galaxy alignment as a source of cosmological information, particularly for the correlation of two cluster shapes.

We found that the effect of weak lensing on clusters is suppressed compared to the case of galaxies. Simulated predictions of the impact of magnification on cluster shapes could establish to a more accurate level the degree of contamination from gravitational lensing to cluster alignments.

\section*{Acknowledgements}

This work is part of the Delta ITP consortium, a program of the Netherlands Organisation for Scientific Research (NWO) that is funded by the Dutch Ministry of Education, Culture and Science (OCW). We thank the {\tt redMaPPer} team and \citet{Vanuitert_2017} for making their data publicly available and Fabian Schmidt and the referee for comments that helped improve our work.

%%%%%%%%%%%%%%%%%%%%%%%%%%%%%%%%%%%%%%%%%%%%%%%%%%
\section*{Data Availability}

Our study relies on third-party public data from the {\tt redMaPPer} team\footnote{\url{http://risa.stanford.edu/redmapper/}} \citep{Rykoff_2014}. For our forecasts, we use v2.1 of the publicly available Core Cosmology Library\footnote{\url{https://github.com/LSSTDESC/CCL}} \citep[{\tt CCL}]{pyccl}.



\bibliographystyle{mnras}
\bibliography{bib} % if your bibtex file is called example.bib

%\begin{thebibliography}{99}
%\end{thebibliography}

%%%%%%%%%%%%%%%%%%%%%%%%%%%%%%%%%%%%%%%%%%%%%%%%%%

%%%%%%%%%%%%%%%%% APPENDICES %%%%%%%%%%%%%%%%%%%%%

%%%%%%%%%%%%%%%%%%%%%%%%%%%%%%%%%%%%%%%%%%%%%%%%%%

% Don't change these lines
\bsp	% typesetting comment
\label{lastpage}
\end{document}